\documentstyle[11pt,newpasp,twoside,epsf]{article}
\def\edcomment#1{\iffalse\marginpar{\raggedright\sl#1\/}\else\relax\fi}
\reversemarginpar

\begin{document}
\title{The Diverse Nature of Intrinsic Absorbers in AGNs}
\author{Fred Hamann}
\affil{University of Florida, Dept. of Astronomy, 211 Bryant Space
Science Center, Gainesville, FL 32653-2055, USA}
\author{Bassem Sabra}
\affil{American University of Science \& Technology, Dept. of Mathematics,
Zahle, Lebanon}

\begin{abstract}
Intrinsic absorbers are significant components of AGN environments 
that provide valuable information and interesting challenges. We 
present a very brief (and biased, and sometimes speculative) overview 
of intrinsic absorbers from the perspective of different 
absorption line classes. We also discuss ways of addressing and 
learning from the ``problem" of partial coverage of the background 
light source, with some examples based on new high-resolution 
rest-frame UV spectra of quasars.
\end{abstract}

\section{Introduction: Why Study Intrinsic Absorbers?}

Let us start by defining ``intrinsic" absorption in terms of gas that
is (or was) part of the overall AGN/host galaxy environment. This
definition excludes only very distant, cosmologically intervening
material, such as intergalactic clouds or unrelated galaxies.
It reminds us that, especially in quasar studies, absorption can
occur in a wide range of environments. The rich variety of intrinsic 
absorbers yields numerous diagnostics of both the AGN phenomenon and
the AGN--host galaxy connection. A short list of reasons for studying 
intrinsic absorption might include the following.

\noindent$\bullet$ Intrinsic absorbers are a fundamental component 
of AGN environments. They are
common in Type I (broad emission line) AGNs (see below), and might be
ubiquitous if the absorbing gas fills only part of the sky as
seen from the central continuum source. In addition, the amounts
of absorbing gas might be enormous --- rivalling or exceeding 
the mass in the broad emission line region.

\noindent$\bullet$ Many intrinsic absorbers are involved in AGN
outflows. The flows are driven by the same 
accretion processes that feed the central super-massive black hole
(SMBH) and fuel other AGN energetics. The need for accreting
matter to expel angular momentum probably means that the wind mass
loss rates, $\dot M_{wind}$, are directly proportional to the mass
accretion rate, $\dot M_{acc}$.

\noindent$\bullet$ The relationship between outflow and accretion 
also
implies that intrinsic absorbers are connected to the basic physics
of SMBH growth and AGN evolution.

\noindent$\bullet$ Intrinsic absorption that occurs far from the AGN
might uniquely measure a variety of regions in the host galaxies, 
such as the interstellar medium, gas streams in galactic halos, or
galactic super-winds driven by starburst activity.

\noindent$\bullet$ The metal abundances in high-redshift intrinsic
absorbers can provide unique constraints on the amount of
star formation and the overall maturity of young galactic or
proto-galactic nuclei.

\noindent$\bullet$ The metal-rich gas expelled by high-redshift
AGNs might be a significant source of metal ``pollution" to the
intergalactic medium at early cosmic times.

In this brief review, we focus on a few issues regarding absorption
line classification, the relationships between classes, and the
implications of partial coverage of the background light
source(s). See also the reviews by Crenshaw, Kraemer, \& George
(2003), Hamann (2000), Hamann \& Ferland (1999), and the ASP
conference series volumes (128 and 255) devoted to AGN mass loss. 

\section{Absorption Line Classes}

AGN absorption lines are classified empirically by the full width at half
minimum (FWHM) of their profiles. The class definitions necessarily
evolve as we encounter new phenomena and assimilate different
measurement schemes. Nonetheless, standardized classes are essential.
The main classes are the ``narrow" absorption lines (NALs), the
``broad" absorption lines (BALs), and a catch-all intermediate class
called the mini-BALs.

BALs are blueshifted from the systemic
(emission line) redshift by as much as $\sim$0.2$c$, and clearly
form in winds form the central engine. Weymann et al. (1991)
introduced a ``BAL-nicity" index to define this class. Hall et
al. (2002) proposed a less restrictive index to include a wider 
range of line widths. We strongly advocate the use of quantitative 
indices, but a reasonable starting point for 
casual conversation is that BALs have continuous absorption
over $>2000$ km/s, with at least some portion of the profile
having a velocity shift $v>2000$ km/s compared to systemic.
BALs are further divided into subclasses according to the degree of
ionization apparent in the lines. ``HiBALs" have nominally SiIV 
$\lambda\lambda$1394,1403 and CIV $\lambda\lambda$1548,1551 
(or perhaps CIII $\lambda$977) as their lowest ionization
lines. ``LoBALs" include lower ionization stages, such as MgII 
$\lambda\lambda$2796,2804. ``FeLoBALs" have more extreme low-ionzation 
regions that produce excited-state absorption in FeII (requiring 
densities $n_e\ga 10^5$ cm$^{-3}$). 

A useful (but physically arbitrary) definition of NALs is that they 
are narrow 
enough {\it not} to blend important UV doublets, e.g., the CIV 
pair with separation $\sim$500 km/s. Thus we require FWHM $< 200$ 
to 300 km/s. NALs with velocity shifts $v<5000$ km/s
from systemic are also call ``associated" absorption lines (AALs)
because of their plausible physical relationship to the AGN. No one
has yet sub-divided AALs into HiAALs and LoAALs analogous to the BALs,
but it would be interesting to compare sub-classes based 
on these properties.

Mini-BALs have FWHMs intermediate between the NALs and BALs. They
appear at the same range of blueshifted
velocities as the BALs, and they also clearly form in AGN
outflows. Examples of high-velocity mini-BALs can be found in Jannuzi
et al. (1996) and Hamann et al. (1997a). For the BALs, see 
Weymann et al. (1991), Reichard et al. (2003), and Hall et
al. (2002). For the AALs, see Foltz et al. (1986) and
Ganguly et al. (1999). See also the reviews cited above.
 
Figure 1 lists the major absorption line classes along with some
properties and speculation that we will describe below. The detection
frequencies given in the figure are our own educated guesses, except
for the quasar AALs (G. Richards, priv. comm.), Seyfert 1 AALs
(Crenshaw et al. 1999), and the HiBALs and LoBALs (Weymann et
al. 1991, Reichard et al. 2003). Note that we list all Seyfert 1
absorbers as NALs, even though some may be considered mini-BALs. 

\begin{figure}
\plotfiddle{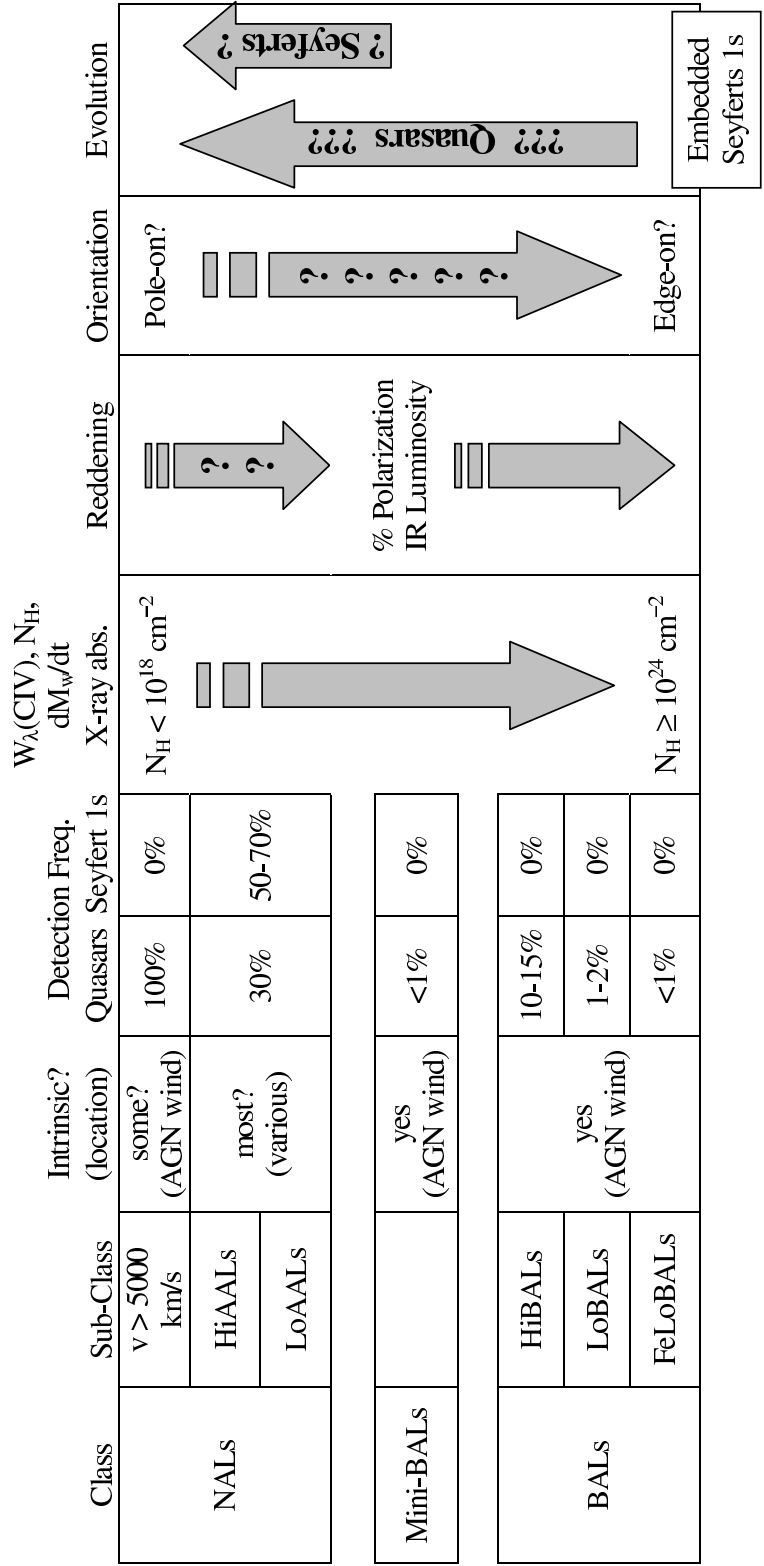}{8.0in}{0.}{92.}{92.}{-245.}{-65.}
%\plotone{hamannf_f1n.ps}
\caption{Absorption line classes and characterictics.}
\end{figure}

\section{Identifying Intrinsic NALs}

Quasar NALs are known to form in a wide range of environments, from
outflows like the BALs to unrelated gas or galaxies at large
(cosmologically significant) distances from the AGN (Weymann et al.
1979). NALs in low-redshift Seyfert 1 galaxies might be almost
exclusively intrinsic, but there is still a range of possible locations
within the global AGN/host galaxy environment. An important
goal, therefore, is to identify the intrinsic NALs and locate the
absorbing regions with respect to the AGNs/host galaxies.
One strategy is to examine the probabilities. For example,
quasars have an over-density of NALs near the emission redshift, 
i.e., more than expected for a random distribution of 
cosmologically intervening material (Weymann
et al. 1979, Foltz et al. 1986). Others have noted that the presence
and/or strengths of NALs correlate with AGN properties such as
luminosity, radio-loudness, and radio lobe orientation (Wills et al.
1995, Richards et al. 1999, Richards 2001). These 
relationships imply that a substantial
fraction of quasar AALs ($>$50\%), and probably some of the highly 
blueshifted ($v> 5000$ km/s) NALs, are intrinsic to quasars.

We can also examine individual absorbers to look for i) absorption line
variability, ii) profiles that are broad and smooth compared 
to thermal speeds, iii) partial line-of-sight 
coverage of the background light source, and iv) high gas densities
based on excited-state absorption lines (see below, also 
Tripp, Lu, \& Savage 1996, Hamann et al. 1997b, 2001,
Barlow \& Sargent 1997, Ganguly et al. 1999, 
Narayanan \& Hamann 2003). These characteristics 
often appear together and suggest an
intrinsic origin because they are most easily understood in terms of 
the dense and dynamic environments near AGNs.

\section{Absorber Trends \& Characteristics}

Brandt, Laor, \& Wills (2000) showed that the 
strengths of intrinsic absorption lines (e.g., $W_{\lambda}$(CIV))
correlate roughly with the amount of X-ray absorption. Moreover, the 
column densities derived for the X-ray absorbers are much larger than 
earlier estimates from the UV lines, to the point where some BAL quasars 
have Thompson thick X-ray absorbers with $N_H \ga 10^{24}$ cm$^{-2}$ 
(Hamann 1998, Gallagher et al. 1999, Mathur et al. 2000, Arav et al. 
2002). In some Seyfert 1 galaxies (e.g., Kaspi et al. 2002) we know that 
the X-ray absorbing gas is outflowing at modest speeds with the UV line
absorbers. It would be more problematic for models invoking 
radiative acceleration if the large X-ray columns in BAL quasars 
turn out to be outflowing at BAL speeds (Hamann 1998). In any
case, we have an expectation that mass loss rates 
increase generally with line strength, as shown in Figure 1.

Other studies suggest that, in very rough terms, the amount of
reddening, the percent polarization, and perhaps the IR luminosities 
are all nominally higher in BAL quasars and may increase toward lower 
ionizations through the BAL class (Weymann et al. 1991, 
Schmidt \& Hines 1999, Brotherton et al. 2002, Hall et al. 2002, 
Richards et al. 2003, Reichard et al. 2003, and refs. therein). 
It is not known if these tendencies extend to the mini-BALs or AALs.
 
\section{Evolution \& Orientation}

AGN evolution is something of a holy grail because we know almost
nothing about it. Unlike stars and galaxies, we have yet to identify
``young" or ``old" AGNs. It has been suggested that the FeLoBAL and
LoBAL quasars represent a young phase because their enormous column
densities, low ionizations, and reddening might be signatures of an
energetic AGN that is still partially enshrouded in its dusty
parental environment (Voit, Weymann, \& Korista 1993, 
Gregg et al. 2000, Brotherton et al. 2002, 
and refs. therein). Perhaps this evolutionary path continues
through the BALs to the mini-BAL and AAL stages as the outflows
gradually weaken (Figure 1). However, orientation probably also plays
an important role. Most models have BAL winds rising 
up out of the accretion disk and flowing close the the disk plane 
(Murray et al. 1995, Proga, Stone, \& Kallman 2000).
Viewing angles close to the disk plane should plausibly lead to
larger absorption column densities, detections of lower ionization
stages, and perhaps increased reddening and polarization (Schmidt 
\& Hines 1999, Elvis 2000). Similar arguments may apply to the 
AALs (Wills et al. 1995). It seems 
likely that orientation and evolution both contribute to differences
between the absorber classes.

One observational difficulty is that AGN lifetimes are short
compared to their host galaxies, so we cannot argue that young/old
AGNs reside in young/old galaxies. In particular, 
AGN luminosities, $L$, are related to
the mass accretion rate onto the central SMBH, $\dot M_{acc}$, by an
efficiency factor, $\eta$. The luminosities are also some fraction,
$\gamma$, of the theoretical Eddingtion limit, $L_{edd}$, such that
\begin{equation}
L \ \approx \ \eta\dot M_{acc}\,c^2 \ \approx \ \gamma L_{edd} \
\approx \ (1.5\times 10^{46})\;\gamma M_8 ~~~ ({\rm ergs/s})
\end{equation}
where $M_8$ is the SMBH mass in units of $10^8$ M$_{\odot}$. If we
adopt standard values of $\eta = 0.1$ and $\gamma = 0.5$, and we
assume the lifetime of the bright AGN phase is the time
needed for the last doubling of the SMBH mass, then we derive a nominal
lifetime of $\sim$$6\times 10^7$ yr (see also Ferrarese 2002 and refs.
therein).

These lifetimes are also short compared to the duration of 
the main quasar epoch, for
example, the $\sim$$5\times 10^9$ yrs elapsed from redshifts 4 to
1. Therefore, we cannot expect to find exclusively young/old AGNs at
high/low redshifts. However, we can use the fact that the AGN
population is growing rapidly at redshifts $z > 3$. At these
redshifts, there are more AGNs being created ``now" than in the past,
so there should be more young AGNs than old ones. Conversely, at
$z\la 1$ the AGN population is in rapid decline and there should be
more old than young AGNs. The strength of this young versus old
signature depends mainly on the slope of the relation between AGN
co-moving number density and time. The signature might be strong 
enough to study age-related trends with redshift in large databases
such as the SDSS. We should also keep in mind that 
the truly youngest AGNs (preceding quasars and having low SMBH
masses) will be found among the least luminous sources, perhaps as 
heavily enshrouded Seyfert 1-like nuclei in dusty starburst 
galaxies (Figure 1).

\section{Column Densities \& Partial Coverage}

Column density is one of the most critical parameters in intrinsic
absorber studies. If we have high
resolution spectra, e.g., of absorption {\it lines}, we can derive
the column density at each velocity shift, $N_v$, from the optical
depth profile, $\tau_v$,
\begin{equation}
N_v \ = \ {{m_e c}\over{\pi e^2 f\lambda}}\; \tau_v
\end{equation}
where $\lambda$ is the line wavelength and $f$ is the oscillator
strength. In principle, this analysis is straightforward. In
practice, there can be observational and other limitations. There is
also a simple question we should consider, which is, what do we mean
by ``the column density" in a particular ion?

If you walk into a room full of astronomers, even AGN astronomers,
and start talking about inhomogeneous absorbing media and partial
coverage of the background light source, you will see some eyes roll
back in their heads while others look longingly for relief at the
clock on the wall. Nonetheless, we keep talking because we believe
partial coverage is a fundamental problem that needs to be addressed
right now in studies of intrinsic absorption. The problem is more
common and probably more insidious than we previously believed.
Continuing progress on the big goals outlined in
\S1 requires a better understanding of the partial coverage issue.

The simplest situation is homogeneous partial coverage (HPC), where
the absorber has a constant optical depth across the projected area
of an extended emission source. The essential geometry is illustrated
in Figure 2a. For simplicity, we assume that the emission source has
a constant intensity, $I_o$, across its projected area. If $\tau_v$
is the optical depth through the absorber at a given velocity shift,
and $C_f(v)$ is the fraction of the emitting area covered by the
absorber (with $0\leq C_f(v) \leq 1$), then the measured intensity is
a spatial average given by,
\begin{equation}
{{I(v)}\over{I_o(v)}} \ = \ 1 - C_f(v) + C_f(v)\, e^{-\tau_v}
\end{equation}
If we measure two lines with a known ratio of optical depths, as in a
doublet, then we can solve for the two unknowns, $C_f(v)$ and
$\tau_v$, using the two equations for the measured intensities
(Barlow \& Sargent 1997, Hamann et al. 1997b, Ganguly et al. 1999).

\begin{figure}
\plotone{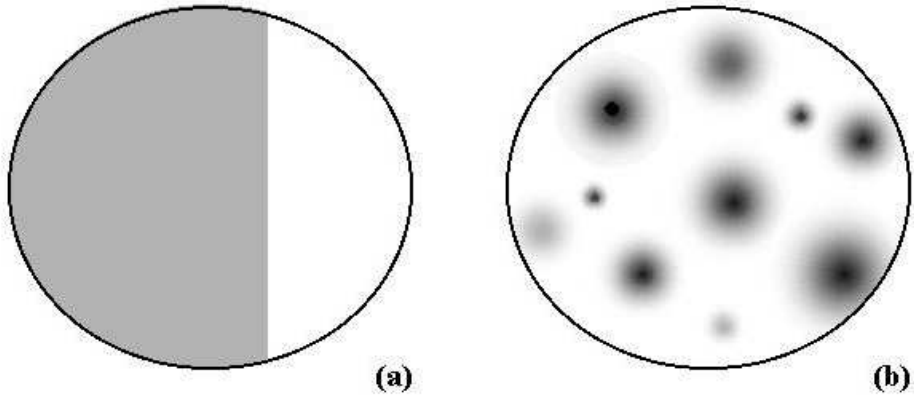}
\caption{Skematic illustrations of homogeneous (a) and inhomogeneous (b)
partial coverage. The white oval areas represent an extended light
source. The shaded regions indicate overlying absorption.}
\end{figure}

There is growing evidence that the reality is {\it
not} like Figure 2a, but more like Figure 2b --- where the
absorber has a range of
optical depths across the projected area of the emission source(s).
We call this situation inhomogeneous partial coverage (IPC, see
deKool et al. 2002, Sabra \& Hamann 2003). The evidence for IPC
absorbers comes from derived coverage fractions that depend on
velocity and, at a given velocity, on the ionization and/or line
strength (i.e., true optical depth, see below, also Barlow et al. 
1997, Hamann et al. 2001, Arav et al. 2002, and refs. therein).
In an IPC cloud
distribution (Figure 2b), strong transitions in abundant ions can
have $\tau_v \ga 1$ over a larger area, leading to larger derived
coverage fractions. Similarly, low ionization species might have
lower coverage fractions if they are concentrated in the smaller,
denser regions.

In general, there is a 2-dimensional optical depth structure,
$\tau_v (x,y)$, in front of an extended emission source with
intensities, $I_o(x,y,v)$, where $x$ and $y$ are the spatial
coordinates. The observed intensity at each velocity shift is the
$I_o$-weighted average of $e^{-\tau_v (x,y)}$,
\begin{equation}
I(v) \ = \ \int\int I_o(x,y,v)\; e^{-\tau_v (x,y)}\; {{dx
dy}\over{A}}
\end{equation}
where $A = \int dx dy$ is the total projected area of the emission
source (deKool et al. 2002, Sabra \& Hamann 2003). deKool et al.
(2002) constructed hypothetical $\tau_v (x,y)$ distributions and
showed how they can affect measured line strengths. They also showed
that any $\tau_v(x,y)$ can be represented by an equivalent
1-dimensional $\tau_v(x)$ without loss of generality. Sabra \& Hamann
(2003) extended the pioneering work of deKool et al. by i)
considering a wider range of $\tau_v(x)$ possibilities, ii) showing
that even in complex IPC situations we can use measured doublets (or
multiplets) to estimate the $\tau_v(x)$ shape, and iii) comparing the
results of IPC and HPC doublet analyses.

Figure 3 shows one hypothetical $\tau_v(x,y)$ distribution composed
of overlapping ``clouds," each with a gaussian optical depth spatial
profile (see Sabra \& Hamann 2003). Also shown is the equivalent
1-dimensional $\tau_v(x)$. We use this optical depth distribution to
calculate intensities for an absorption line doublet. As in the HPC
analysis, the doublet intensities provide two equations from which we
can derive two unknowns. But here we adopt a simple functional form
for the optical depths, such as the power law $\tau_v (x) =
\tau_{max}\; x^a$, and then solve for $\tau_{max}$ and the ``shape"
parameter $a$. The results are shown in Figure 3. The power law
provides a close but imperfect match to the input $\tau_v(x)$ because
the input was {\it not} a
power law. To do better, e.g., in the analysis of real AGN lines, we
need theoretical guidance on what functional forms of $\tau_v(x)$
are appropriate. It would also help to measure sets of $\geq$3
lines with known optical depth ratios, e.g., in multiplets, 
to obtain more constraints on $\tau_v(x)$.

\begin{figure}
\plottwo{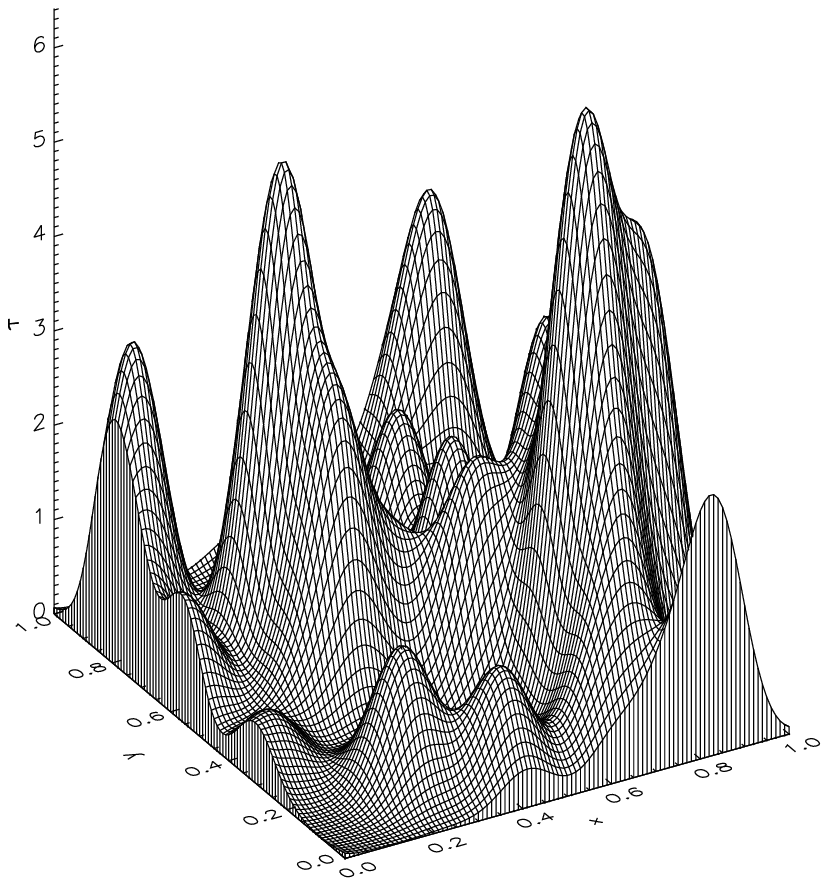}{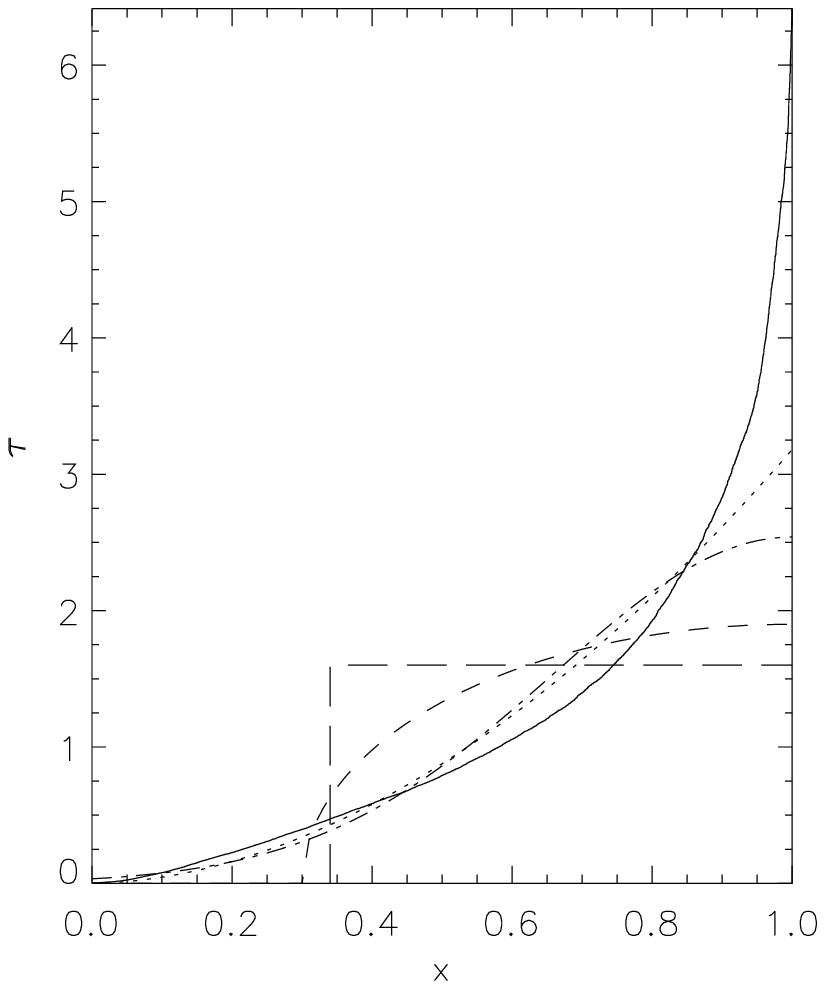}
\caption{Left panel: A synthetic $\tau_v(x,y)$ distribution.
Right panel: The equivalent $\tau_v(x)$ (solid line) plus the best
solutions from a doublet analysis assuming a power law $\tau_v(x) =
\tau_{max}\; x^a$ (dotted line) and several other assumed $\tau_v(x)$
shapes, including a step function HPC scenario (long-dash line). From
Sabra \& Hamann (2003).
}
\end{figure}

Finally, we return to the question of what we mean by ``the
optical depth" or ``the column density" in a given feature. Are there
single values of these parameters that are meaningful in an IPC
environment, e.g., that we can use in a simple ionization/abundance
analysis? The answer is probably yes. The input $\tau_v(x,y)$
distribution in Figure 3 has a spatially averaged optical depth of
$\tau^{avg}_{IPC}\approx 1.2$. The HPC doublet analysis applied to
the line intensities emerging through that distribution
yields $\tau^{avg}_{HPC}
\approx 1.1$. Sabra \& Hamann (2003) performed many such experiments.
We found that $\tau^{avg}_{IPC}/\tau^{avg}_{HPC}\approx 1$ to
within $\sim$30\% or so, as long as the input $\tau_v(x,y)$ does not
contain spatially narrow ``spikes." Such spikes (very large optical
depths over a small coverage area) can profoundly affect the average
optical depth while having little impact on the observed intensities.
Therefore, even with potentially complex IPC absorbers,
we can use spatially averaged optical depths derived from the
HPC analysis to obtain spatially averaged estimates of the
column densities, ionization, abundances, etc.

\section{Applications to Real Data}

We obtained (with our collaborators T. Barlow, E.M. Burbidge, \& V.
Junkkarinen) high-resolution spectra of a sample of AAL quasars using
the HIRES spectrograph at the Keck 10 m telescope. The quasars have
redshifts of order 2, allowing us to measure absorption lines in the
rest-frame UV. Figure 4 shows line profiles for several doublets
in two of these quasars. The HPC doublet analysis shows that
these lines are optically thick at all velocities, while
velocity-, ion-, and line-dependent
coverage fractions determine the line strengths and
profiles (see also Barlow et al. 1997, Arav et al. 2002, and refs. 
therein). Each individual line (or velocity component) 
probably does {\it not} correspond to a single entity we might call 
a ``cloud" in the absorber. These features 
will provide only lower limits on the ionic column 
densities. This situation seems to be common for 
intrinsic quasar NALs with FWHMs $\ga$ 100 km/s. 
However, narrower quasar NALs appear to have typically $C_f(v)\approx 1$, 
with each distinct velocity component corresponding to a distinct cloud. 
This simpler circumstance can yield more reliable column densities 
and abundances (we find typically $Z\ga$ Z$_{\odot}$).

\begin{figure}
\plottwo{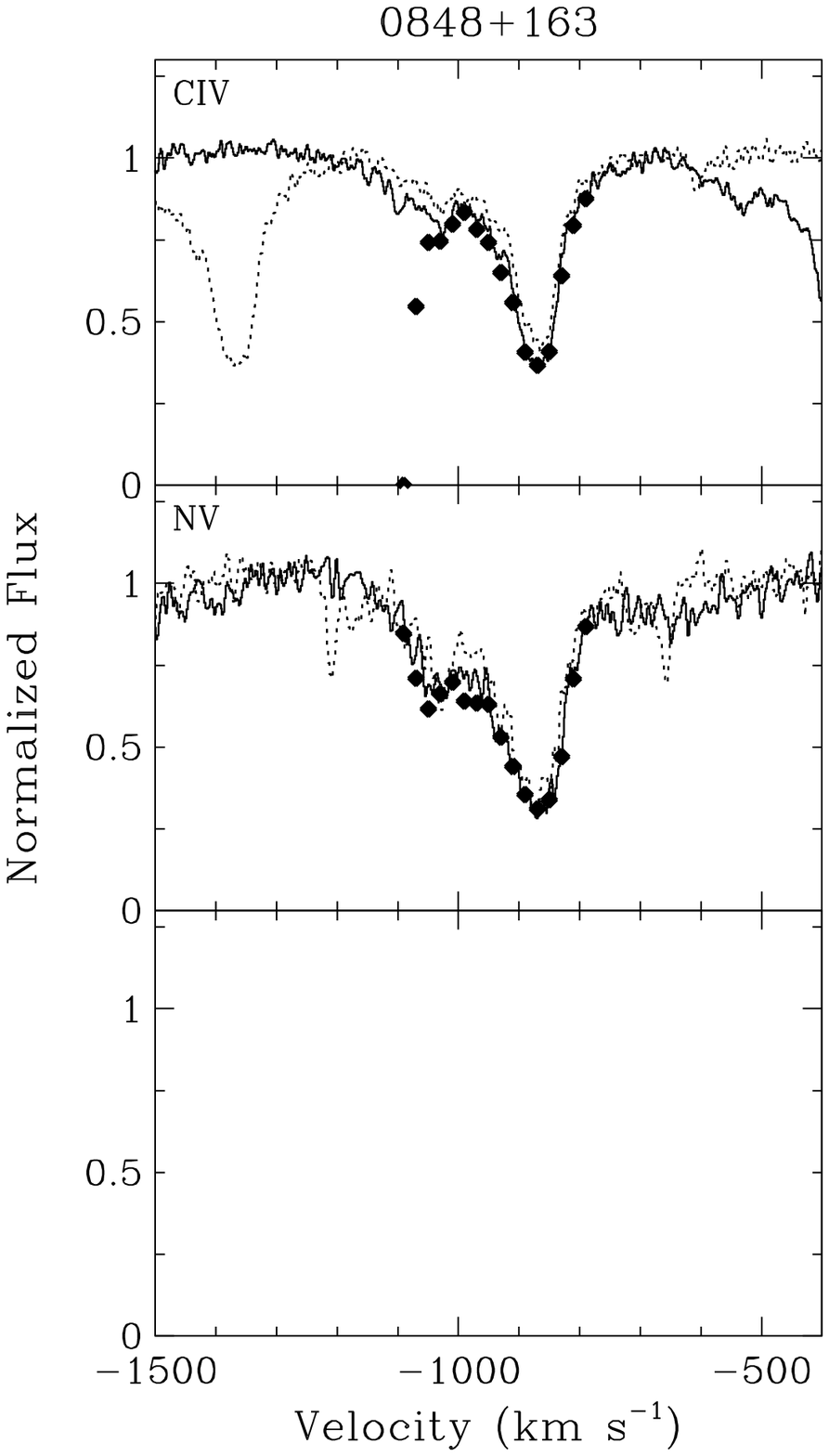}{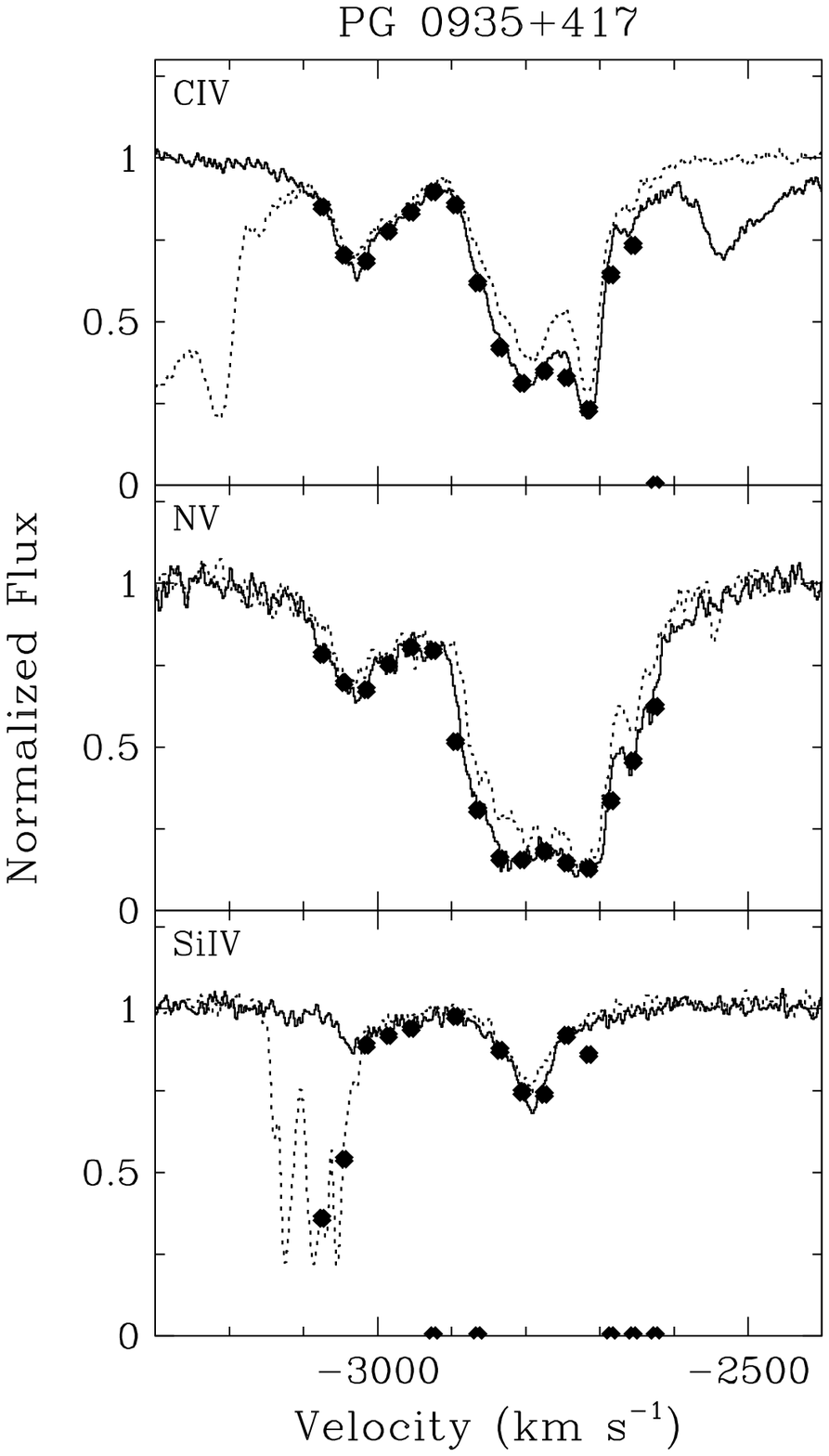}
\caption{AAL doublet profiles in CIV, NV
$\lambda\lambda$1238,1243, and SiIV in two
quasars at redshift $z\sim 2$. The velocities are relative to the broad
emission lines. The solid/dotted curves represent the
short/long wavelength doublet components, which all have inherently 2/1
optical depth ratios. Derived values of $1-C_f(v)$ \ (large dots)
indicate that the lines are optically thick ($\tau_v > 3$) and
velocity-dependent coverage determines each line's strength and profile.
(Noise in the data and overlap with unrelated lines sometimes leads to
spuriously low $1-C_f(v)$ results, e.g., in SiIV in PG~0935+417.)}
\end{figure}

\end{document}